\documentclass[floatfix,reprint,amsmath,amssymb,aps,superscriptaddress]{revtex4-2}

\usepackage[mathscr]{euscript}
 \let\mathscr\relax
\usepackage[scr]{rsfso}

\usepackage{float}
\usepackage{graphicx}
\usepackage{xcolor}
\usepackage{xparse}
\usepackage[caption=false]{subfig}
\usepackage{dcolumn}
\usepackage{bm}
\usepackage{amsmath}
\usepackage{physics}
\usepackage{array}
\usepackage{bbold}
\usepackage{commath}
\usepackage{braket}
\usepackage{dsfont}
\usepackage{tikz}
  
\ExplSyntaxOn

\keys_define:nn { miguel/label }
 {
  label   .tl_set:N = \l_miguel_label_tl,
  unknown .code:n   = \clist_put_right:Nx \l_miguel_label_clist
                       { \l_keys_key_tl = \exp_not:n { #1 } }
 }
\clist_new:N \l_miguel_label_clist
\box_new:N \l_miguel_label_box
\box_new:N \l_miguel_label_image_box

\NewDocumentCommand{\xincludegraphics}{O{}m}
 {
  \tl_clear:N \l_miguel_label_tl
  \clist_clear:N \l_miguel_label_clist
  \keys_set:nn { miguel/label } { #1 }
  \tl_if_empty:NTF \l_miguel_label_tl
   {
    \miguel_includegraphics:Vn \l_miguel_label_clist { #2 }
   }
   {
    \hbox_set:Nn \l_miguel_label_image_box
     {
      \miguel_includegraphics:Vn \l_miguel_label_clist { #2 }
     }
    \hbox_set:Nn \l_miguel_label_box
     {
      \skip_horizontal:n { 3pt }
      \fcolorbox{white}{white}{\footnotesize \tl_use:N \l_miguel_label_tl}
     }
    \leavevmode
    \box_use:N \l_miguel_label_image_box
    \skip_horizontal:n { -\box_wd:N \l_miguel_label_image_box }
    \hbox_overlap_right:n
     {
      \box_move_up:nn
       {
        \box_ht:N \l_miguel_label_image_box - 
        \box_ht:N \l_miguel_label_box - 3pt
       }
       { \box_use_drop:N \l_miguel_label_box }
     }
    \skip_horizontal:n { \box_wd:N \l_miguel_label_image_box }
   }
 }

\cs_new_protected:Nn \miguel_includegraphics:nn
 {
  \includegraphics[#1]{#2}
 }
\cs_generate_variant:Nn \miguel_includegraphics:nn { V }

\ExplSyntaxOff

\DeclareMathOperator*{\maxi}{max}
\DeclareMathOperator*{\mini}{min}

\begin{document}
\title{Entropic Accord: A new measure in the quantum correlation hierarchy}

\author{Biveen Shajilal}
\email{biveen.shajilal@anu.edu.au}
\affiliation{Centre for Quantum Computation and Communication Technology, Research School of Engineering,
The Australian National University, Canberra, ACT 2601, Australia}
\affiliation{Centre for Quantum Computation and Communication Technology, Department of Quantum Science, The Australian National University, Canberra ACT 2601, Australia}

\author{Elanor Huntington}
\affiliation{Centre for Quantum Computation and Communication Technology, Research School of Engineering,
The Australian National University, Canberra, ACT 2601, Australia}

\author{Ping Koy Lam}
\affiliation{Centre for Quantum Computation and Communication Technology, Department of Quantum Science, The Australian National University, Canberra ACT 2601, Australia}
\affiliation{School of Physical and Mathematical Sciences, Nanyang Technological University, Singapore 639673, Republic of Singapore}

\author{Syed Assad}
\affiliation{Centre for Quantum Computation and Communication Technology, Department of Quantum Science, The Australian National University, Canberra ACT 2601, Australia}
\affiliation{School of Physical and Mathematical Sciences, Nanyang Technological University, Singapore 639673, Republic of Singapore}

\date{\today}

\pacs{Valid PACS appear here}

\maketitle

\section*{Abstract}

Quantum correlation often refers to correlations exhibited by two or more local subsystems under a suitable measurement. These correlations are beyond the framework of classical statistics and the associated classical probability distribution. Quantum entanglement is the most well known of such correlations and plays an important role in quantum information theory. However, there exist non-entangled states that still possess quantum correlations which cannot be described by classical statistics. One such measure that captures these non-classical correlations is \emph{discord}. Here we introduce a new measure of quantum correlations which we call \emph{entropic accord} that fits between entanglement and discord. It is defined as the optimised minimax mutual information of the outcome of the projective measurements between two parties. We show a strict hierarchy exists between entanglement, entropic accord and discord for two-qubit states. We study two-qubit states which shows the relationship between the three entropic quantities. In addition to revealing a class of correlations that are distinct from discord and entanglement, the entropic accord measure can be inherently more intuitive in certain contexts.

\section{\label{level1}Introduction}

Bipartite states can have correlations that cannot be described by classical joint probability distributions. To some degree these non-classical correlations can be explained using entanglement and more generally using quantum discord. Quantum entanglement plays a significant role in the developmental \emph{timeline of quantum mechanics}. The term entanglement was originally coined by Schrodinger to explain the statistical correlations between subsystems of a joint-quantum system~\cite{schrodinger}. This "spooky-action" was explicitly pointed out in the seminal publication of Einstein, Podolsky, and Rosen, which suggested that under reasonable assumptions quantum theory must be incomplete~\cite{einstein1935can}. This would imply the existence of a global state that cannot be written as a product of the individual states. Quantum entanglement is intriguing because it challenges the experience and intuition on how the macroscopic world work. Thirty years later, Bell proposed a thought experiment to test the predictions of a local hidden variable theory~\cite{freistadt1957causal,kochen1975problem} and quantum theory~\cite{bell1964einstein}. This led to the proposal of the Clauser, Horne, Shimony, and Holt (CHSH) inequality which holds if \emph{Local Hidden Variable} theory is true ; whereas it could be violated if nature follows the quantum mechanical description~\cite{clauser1969proposed}. The simplicity of the experiment led to its immediate demonstration by Freedman and Clauser. They experimentally showed the violation of the CHSH inequality, thus validating the modern quantum theory~\cite{FreedmanClauser}. Aspect~\textit{et al.}~and several others followed up with improved experiments including the more recent loophole free demonstrations~\cite{Aspect1,Aspect2,Aspect3,entreview,hensen2015loophole}.

Quantum entanglement is a pivotal resource in quantum information. In most quantum computing protocols, the lack of entanglement in the system would mean that the computational tasks could be emulated on classical platforms~\cite{jozsa2003role}. Entanglement is thus a distinguishing factor between the two computing platforms.  A famous example is the presence of entanglement between the register qubits in Shor's algorithm~\cite{shor1994algorithms}.

It is tempting to assume that entanglement is the requisite resource that delivers quantum advantage for all quantum protocols. This is indeed true for systems with pure states but not necessarily the case for mixed states systems. Deterministic Quantum Computation with One Bit (DQC1) is an example of such a mixed state quantum protocol that does not rely on entanglement~\cite{knill1998power}. This algorithm perform some key computational tasks better than any known classical equivalent algorithm with maximally mixed state that has little to no entanglement. These mixed states protocols are closer to experimental conditions that are never ideal. In this context, the role of entanglement as a fundamental resource is not directly obvious. This conundrum has sparked new research to search for other useful measures to more accurately quantify the requisite quantum correlations for quantum information.

Datta \textit{et al.}~\cite{datta2008quantum} showed that the states involved in DQC1 exhibited the correlation quantum discord  proposed by Zurek and Vedral~\cite{zurek2000,zurek2001,Vedral}. Discord is the difference between quantum mutual information and classical mutual information. Quantum mutual information is identical to classical mutual information when the involved systems are classically correlated and differs when the systems involved are quantum. Datta~\textit{et al.}~showed that the non-vanishing discord present was responsible for the quantum advantage~\cite{datta2008quantum}. Quantum discord coincides with entanglement for pure states and differs only for mixed states. Eastin~\textit{et al.}~followed up Datta's work by showing that mixed state computation with zero discord could be classically simulated~\cite{eastin2010simulating}. These findings incentivised further study into mixed state correlations outside the prevalent concept of entanglement.

\begin{figure}[htp]
        \includegraphics[width=0.42\textwidth]
        {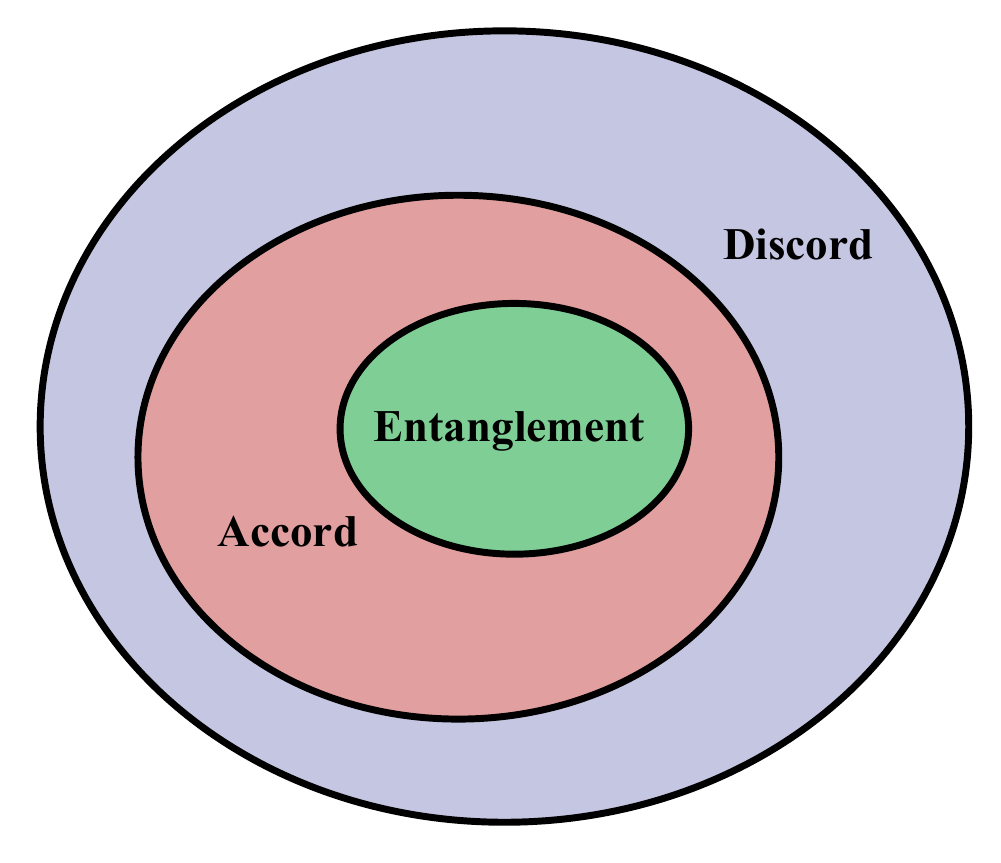}
        \caption{\label{hierarchy}Schematic representation of the correlation hierarchy. All entangled states have a finite value of accord and all states with accord have a finite value of discord whereas the contrary statements are not true.}
\end{figure}

Recently a new measure, \textit{accord}, was proposed by Szasz \cite{PhysRevA.99.062313} which has the interesting trait of approximately lying between discord and entanglement. One motivation for this new measure was to find a correlation measure that is stronger than discord but not as strong as entanglement. This would mean the correlation hierarchy can be pictorially represented as shown in Fig.~\ref{hierarchy}.

The rudimentary idea behind accord is to measure correlations through a sequence of adversarial actions. This is unlike entanglement where the correlations represent cooperative subsystems. In the thought experiment which establishes the accord measure, Alice tries her best to perform adversarial measurements against Bob's effort to maximise the mutual information. This is unconventional as we often associate cooperation to correlations. The thought experiment proposed by Szasz define the correlation measure as the optimal measurement coincidence probability (OMCP) when the two parties are involved in a minimax game. In essence OMCP accord is the rescaled probability of Alice correctly guessing Bob's measurements despite Bob's efforts to work against it. Accord satisfies the most important qualities of a quantum measure of being non-negative and equal to zero for classical states. However, Szasz’s measure results in some entangled states having zero accord. Then the desired hierarchy shown in Fig.~\ref{hierarchy} is not necessarily true but one that comes with certain exceptions. Still the thought that there exists correlations that have certain unique characteristics in comparison to the others in the more generalised correlation family generates much interest in formulating new quantum resource measures. Quantum games in quantum mechanics have been of huge interest, mostly used as a tool to explore the interesting field of quantum information. They have been investigated in the context of quantum state estimation and cloning~\cite{lee2003game}, for developing novel quantum algorithms~\cite{meyer1999quantum} and show potential for further research~\cite{lee2002let}. An optical analogue~\cite{englert2001universal} of the mean king's problem~\cite{vaidman1987ascertain,englert2001mean} shows potential application in quantum communication protocols~\cite{beige2002secure}. These are indications of how diversified quantum games can be. Szasz's take on using an  adversarial quantum game to define a new correlation is an intriguing direction with potentially interesting implications.

In this paper, we present a new measure of quantum correlations which we call \emph{quantum entropic accord} (EA) which is the mutual information minimaxed over all possible local projective measurements by the two parties playing an adversarial game. The measure is inspired by quantum discord which is based of mutual information and the interesting game proposed by Szasz. We looked into the possibility of a correlation measure based on mutual information of similar characteristics. Optimised mutual information between both parties is now positive even for the class of states that initially troubled OMCP accord with zero value while being entangled.

This paper is organised as follows. In section II~(A), we describe the thought experiment and introduce EA as a measure. In section II~(B), we prove the correlation hierarchy for two-qubit systems and evaluate numerically the EA for several classes of states including pure states, Bell-diagonal states and classical states. The reason behind choosing these specific classes of states is primarily that of experimental relevance as well as the availability of well formulated correlation measures. This enables us to readily compare the EA with an entanglement measure and discord. We also look at arbitrary two qubit states for comparing EA with other measures of correlations. Finally, we conclude with a summary and discussions in section III.

\section{Results}
\subsection{\label{level3}Definition of Entropic Accord}
\begin{figure}[htp]
        \center{\includegraphics[width=0.475\textwidth]
        {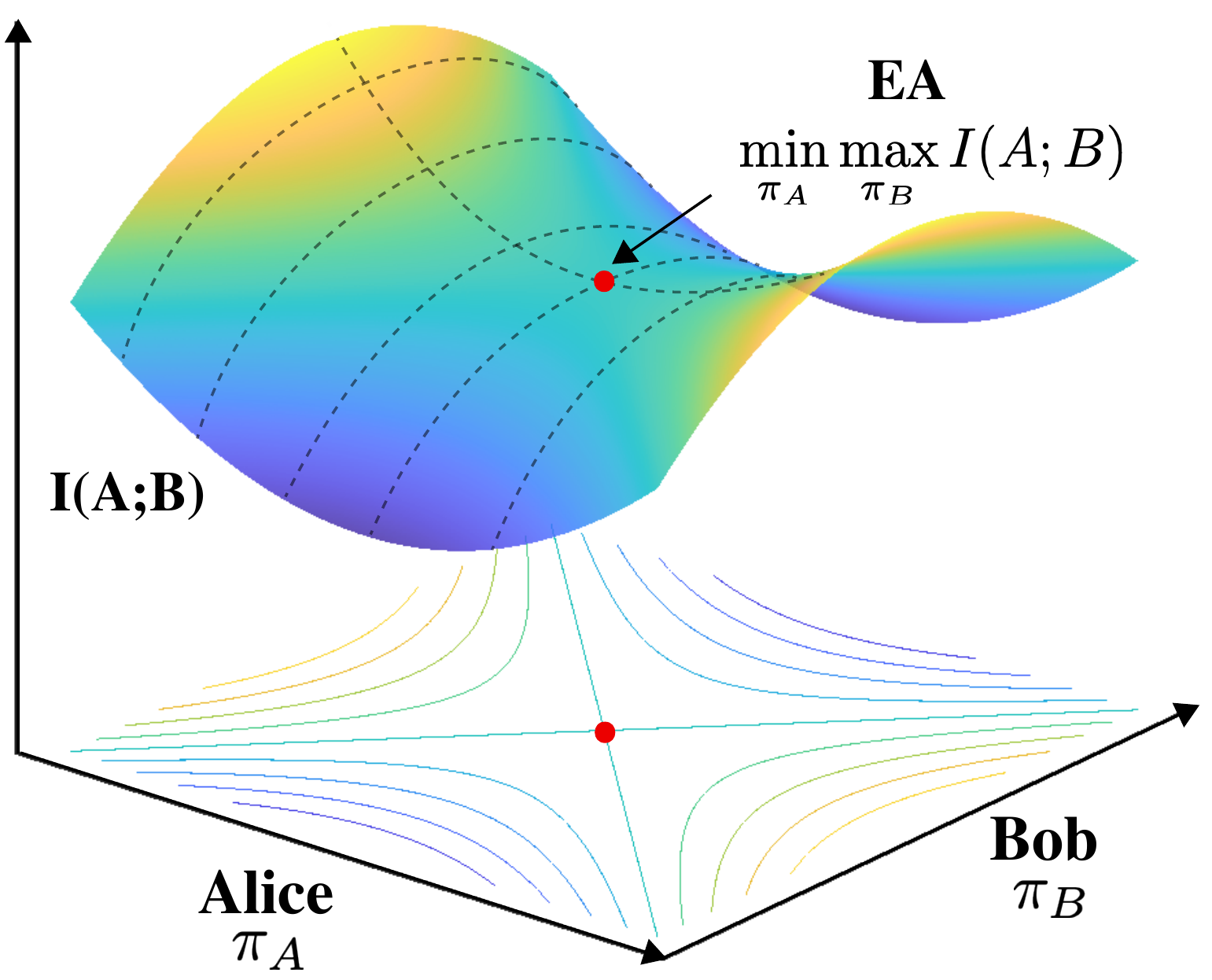}}
        \caption{\label{thegame} Pictorial representation of the game Alice and Bob plays to establish entropic accord (EA). Alice performs a projective measurement to minimize the information that she would potentially share with Bob. With this knowledge, Bob subsequently makes his projective measurement to maximize the mutual information. The saddle point of the mutual information quantifies the EA.}
\end{figure}

The thought experiment which defines the new correlation measure involves a minimax game between Alice and Bob. One particular example for a minimax game would be chess. In a perfect game of chess with Alice making the next move, her best strategy would be to choose a move that maximises her chances of winning with the expectation that Bob will respond with a move to minimise her chance of winning. In other words, Alice must act in a way anticipating Bob will choose the best move. In the context of quantum information, the minimax algorithm can be formalised as follows. Consider Alice (A) and Bob (B) spatially separated sharing a joint quantum state $\rho$. They are capable of performing projective measurements. Alice wants to minimize the correlation and Bob wants to maximize it in the following fashion,
\begin{enumerate}
\item Alice choose a projective measurement $\pi_A$ to minimize the mutual information given her knowledge of $\rho$ and shares the information of her measurement with Bob.
\item Bob choose a projective measurement $\pi_B$ to maximize the mutual information given his knowledge of Alice's measurement $\pi_A$ and the shared quantum state $\rho$.
\item For the $\rho$, $\pi_A$ and $\pi_B$ from the previous steps, they do their respective projective measurements and the mutual information quantifies the correlation of their measurements.
\end{enumerate}
We define the new correlation measure as:
\begin{equation}
\text{Entropic Accord} = \mini_{\pi_{A}}\maxi_{\pi_{B}} I(A;B), \label{perf}
\end{equation}
where $I(A;B)$ is the classical mutual information calculated from the joint probability $P(A,B)$ and the marginal probabilities, $P(A)$ and $P(B)$,
\begin{equation}
I(A;B) = \sum_{A} \sum_{B} P(A,B) \text{log} \frac{P(A,B)}{P(A)P(B)}
\end{equation}
Here the minimisation and maximisation are over projective measurements $\pi_{A}$ and $\pi_{B}$. The scheme is pictorially represented in Fig.~\ref{thegame},
which provides an intuitive picture of the game where Alice's intention is to minimise the mutual information through her measurements. She starts with the knowledge of what state is shared between her and Bob. However Bob's intention is the opposite, with his knowledge of Alice's measurement, he tries to maximise the mutual information. This degree of correlation arises from their correlated subspaces as a result of the procedure mentioned above or the game both parties play. Intuitively, EA is the smallest information Bob can force without knowing Alice's measurement also the largest information Alice can guarantee if she knows Bob's measurement.

We first check that EA satisfies the requisites for being a good correlation measure. Brodutch~\textit{et. al}~\cite{brodutch2011criteria} proposes three conditions for this: 
\begin{enumerate}
	\item The measure should be nonnegative. 
	\item The measure is invariant under local unitary transformations.
	\item The measure is zero for classically correlated states.
\end{enumerate} 
In our favour, by definition EA is invariant under local unitary transformations and since mutual information is nonnegative, the first two conditions are met. The final criterion to check is whether this measure is zero for classically correlated states. Classical states are characterised by completely diagonal reduced density matrices. 
\begin{equation}
\rho_{c} = \sum_{m,n} \ket{m,n} c_{m,n} \bra{m,n},
\end{equation}
where $\{\ket{n}\}$ forms an orthogonal basis. Assume Alice measures in a basis that is mutually unbiased to \{$\ket{m}$\}. Then Bob's reduced state $\rho_{B|A}(a=m)$ does not depend on $m$ which implies $I_{AB} = 0$, which means EA is zero for classical states.

\subsection{\label{level2}Entropic Accord: Two-qubit systems}
As pointed out earlier, the usefulness of a quantum state as a resource for quantum information protocol is typically quantified by the amount of entanglement present. Discord is beyond the framework of this school of thought as some states with zero entanglement and non-zero discord still shows benefits in quantum protocols \cite{bradshaw2017overarching,datta2008quantum}. Discord is the difference between the quantum mutual information and the classical mutual information between Alice and Bob where the classical measurements occurs at Bob's side \cite{datta2009signatures}. Computing quantum discord for a general case is an NP-complete problem. However for two-qubit systems it can be calculated easily and for certain classes of states, such as Bell-diagonal states where we can calculate discord analytically. In the following subsections we will compare EA with entanglement and discord in bipartite qubit systems. 

\subsubsection{Proof of the correlation hierarchy}
As shown earlier EA fulfils the three conditions for being a correlation measure, however it is important to look whether a strict hierarchy holds true for EA unlike in the case of the OMCP accord. For a general two-qubit state, we can show that zero discord $\Rightarrow$ zero EA $\Rightarrow$ zero entanglement. A state has zero discord if and only if it can be written as,
\begin{equation}
\rho_{AB} = \sum_{j} \rho_{B,j} \otimes \ket{e_{j}}\bra{e_{j}}.
\end{equation}
When Alice chooses to measure in a basis that is mutually unbiased to $\ket{e_j}$, then Bob's reduced state will always be the same. Consequently the mutual information is zero and therefore EA is zero.\\
Similarly, a two-qubit state has zero EA if and only if the state $\rho_{AB}$ can be written as,
\begin{equation}
\begin{split}
\rho_{AB} & = \ket{0}\bra{0} \otimes \rho_{B}  + \ket{0}\bra{1} \otimes \phi_{B}   + \ket{1}\bra{0} \otimes \phi_{B}^{\dagger} \\ 
& + \ket{1}\bra{1} \otimes \rho_{B},
\end{split}
\end{equation}
where $\rho_B$ represents the conditional state of Bob and $\phi_{B}$ is an arbitrary operator such that $\rho_{AB} \geq 0$. This state has zero EA because if Alice measures in the computational basis, Bob's state will always be $\rho_B$ regardless of Alice's outcome. If a state cannot be written in this form, then Bob's conditional state will be different and the EA will be nonzero. As every density matrix $\rho_{AB}$ is positive semidefinite, states of this form have positive partial transpose.
i.e.,
\begin{equation}
\begin{bmatrix}
    \rho_{B} & \phi_{B}  \\
    \phi_{B}^{\dagger} & \rho_{B}
  \end{bmatrix} \geq 0 \Leftrightarrow \begin{bmatrix}
    \rho_{B} & \phi_{B}^{\dagger}  \\
    \phi_{B} & \rho_{B}
  \end{bmatrix} \geq 0 ,
\end{equation}
\begin{equation}
\rho_{AB} \geq 0 \Leftrightarrow \rho_{AB}^{T_{A}} \geq 0
\end{equation}
This is a sufficient condition for two-qubit systems to be separable and consequently leads to zero entanglement \cite{peres1996separability,horodecki2001separability}.

\begin{figure}[t]
        \center{\includegraphics[width=0.5\textwidth]
        {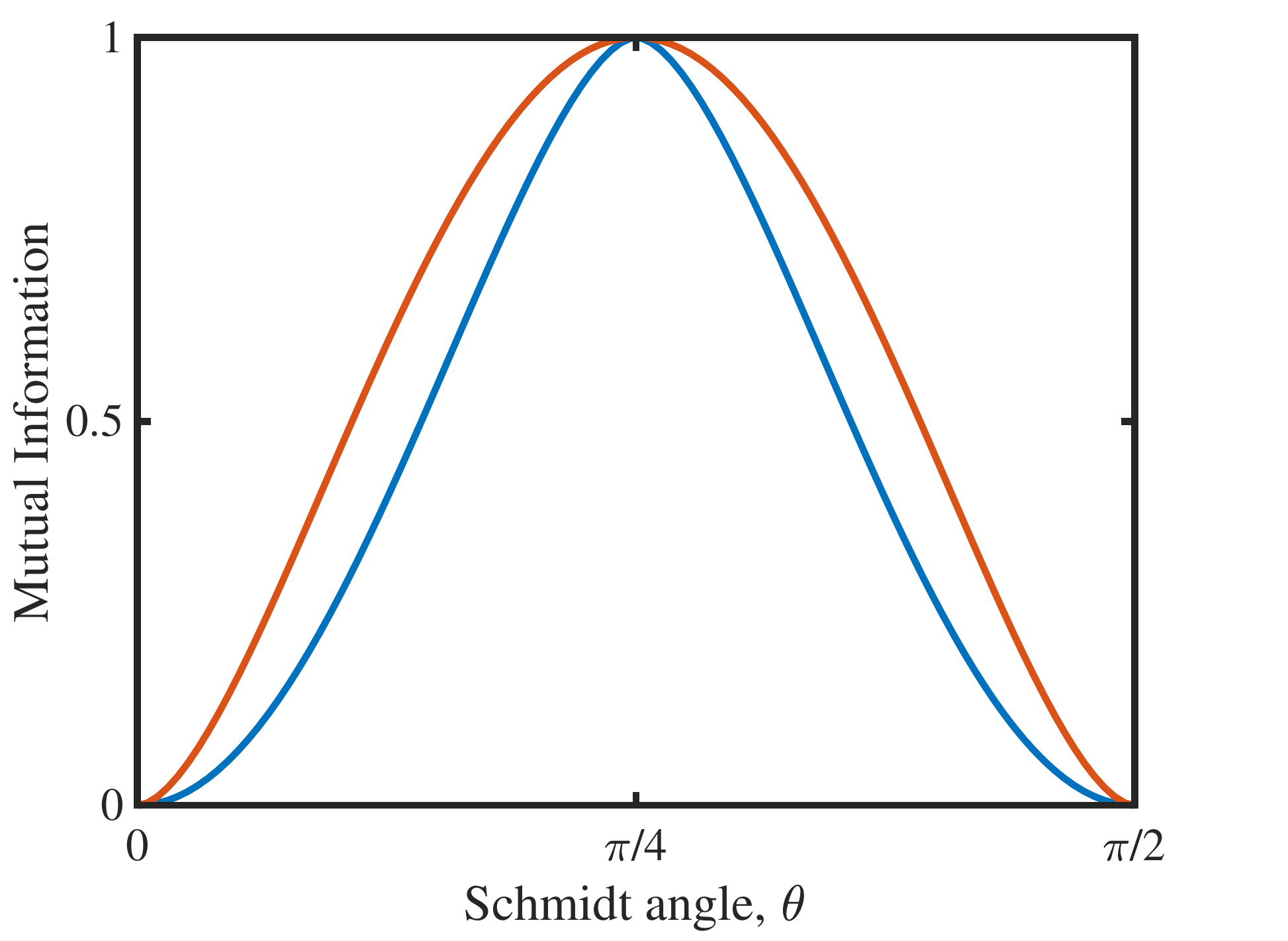}}
        \caption{\label{upper} Mutual information as a function of the Schmidt angle $\theta$. The red curve represents the entropy of entanglement and the blue curve represents the upper bound to EA when Alice choose to measure in the $\sigma_{x}$ basis for pure states. For pure states, quantum discord reduces to entanglement.}
\end{figure}

\subsubsection{Pure states and the addition of white noise}

\begin{figure*}[htp]
\centering
 \subfloat{%
    \xincludegraphics[width=.33\textwidth,label=a)]{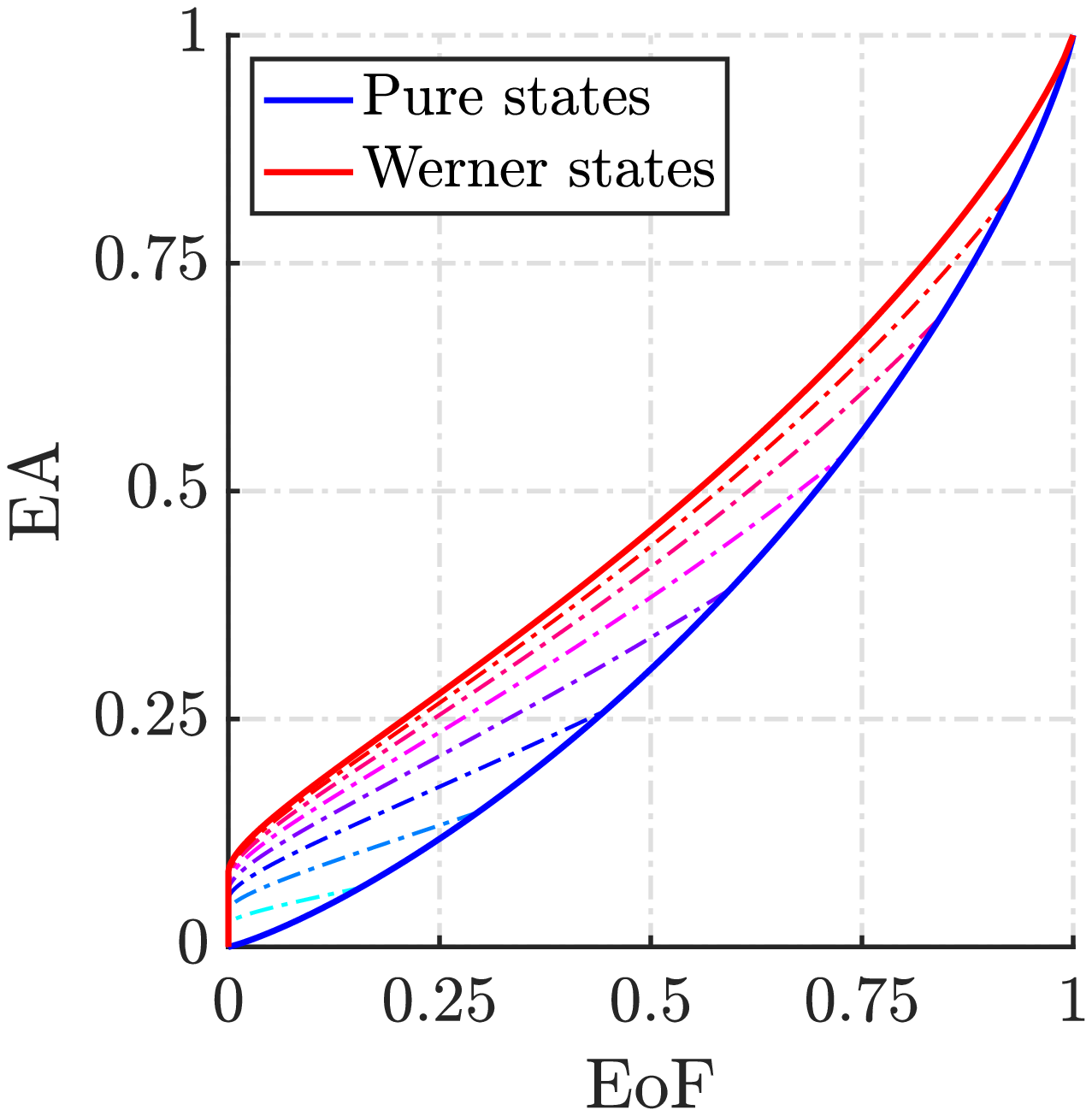}}\hfill
  \subfloat{%
   \xincludegraphics[width=.33\textwidth,label=b)]{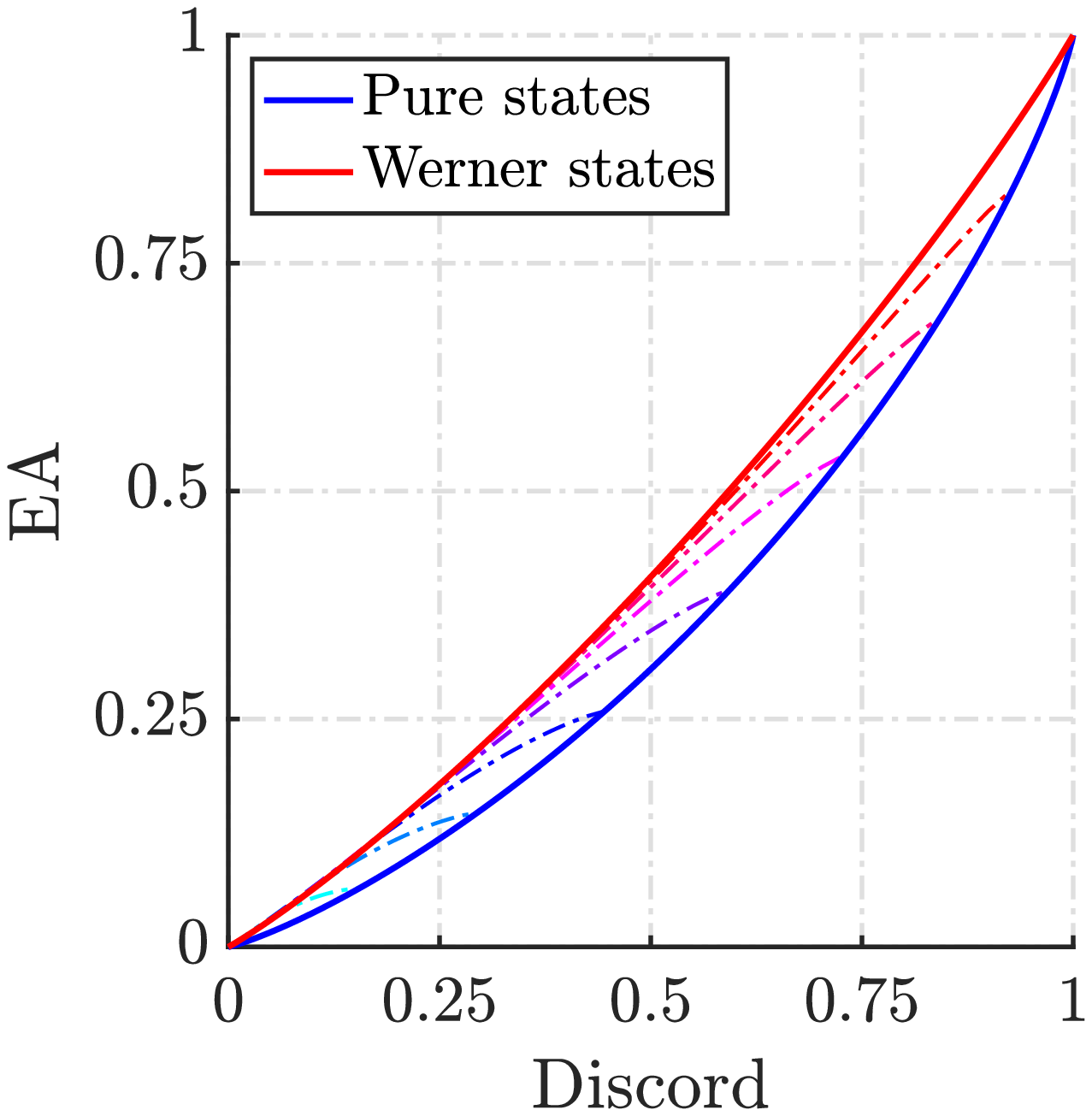}}\hfill
  \subfloat{%
   \xincludegraphics[width=.33\textwidth,label=c)]{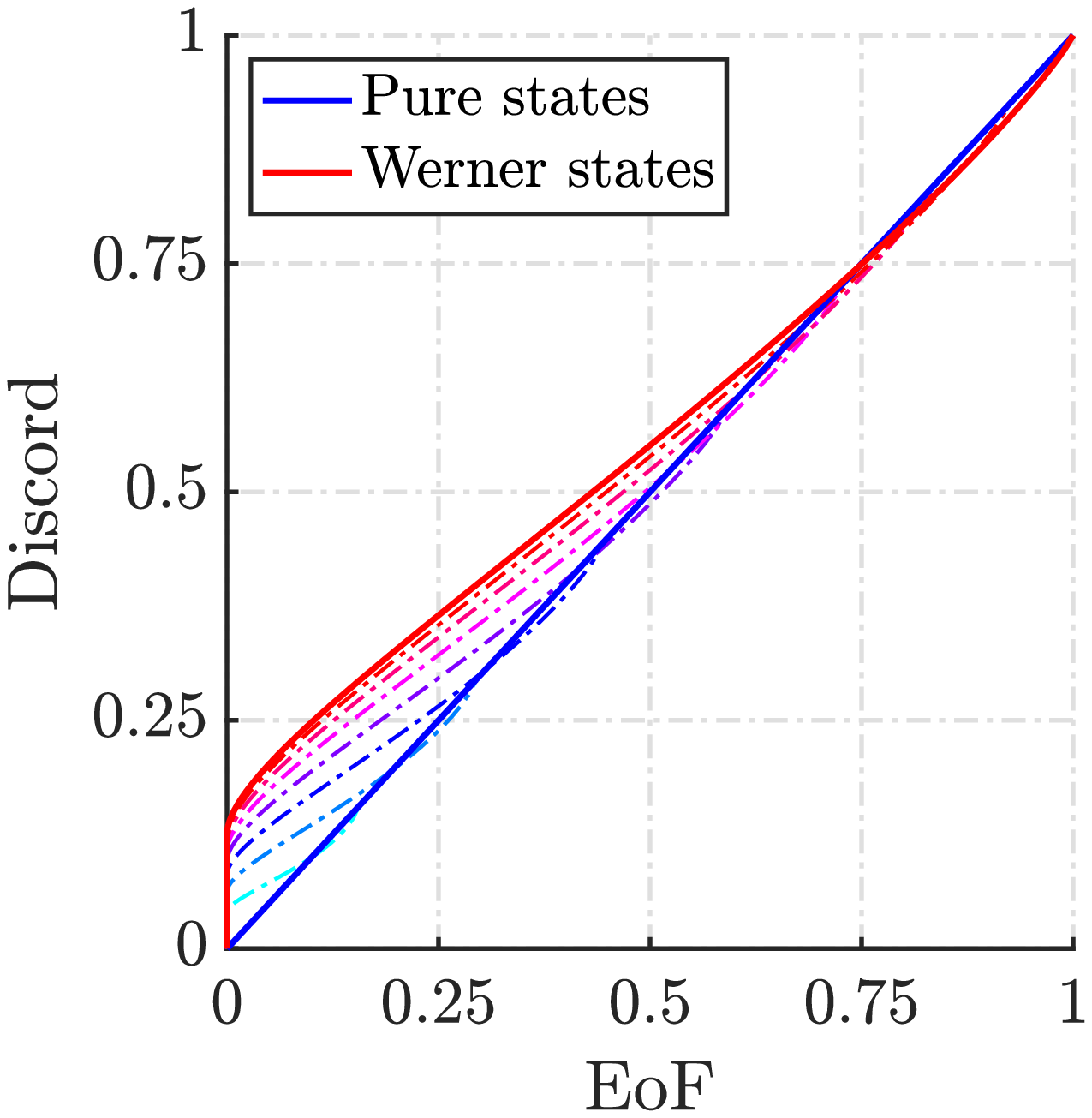}}
 \caption{Comparison of EA with entanglement of formation~(EoF) and discord for pure states with white noise. Each colour represents the group of states which is a mixture of noise and pure state parametrised by the Schmidt angle $\theta$. Each line represents how the correlations evolve for pure state as we introduce noise. From (c) it is evident that discord asymptotes to entanglement for pure states whereas EA does not. EA does not asymptote to both the measures for the class of states under comparison. For a given EoF and discord, Werner states have the maximum EA. Werner states are states of the form $(1-e)\ket{\psi_-}\bra{\psi_-} + e\mathbb{1}/4$ with $\psi_-$ being the Bell singlet. Werner states are bipartite quantum states which are invariant under all unitary transformations. }\label{pure_state_noise_same_pure_states} 
\end{figure*}

In case of pure states, quantum discord reduces to entanglement. As we shall show, this is not the case with EA. For pure states, entanglement is quantified using the von Neumann entropy of the reduced state of each party. Any bipartite pure state can be brought to Schmidt form by local operations \cite{sciara2017universality,zyczkowski2006geometry},
\begin{equation}
\ket{\Psi} = \ket{00} \cos \theta + \ket{11} \sin \theta,\label{pure_eqn}
\end{equation}
where $\theta$ is the Schmidt angle. In the context of a game that establishes a measure, if the two parties are cooperating with each other and decide to measure in the $\sigma_z$ basis, the entropy of entanglement of this state would correspond to the mutual information between Alice and Bob. The outcomes of these measurements will look like,
\begin{center}
\begin{tabular}{| c | c  c |}
\hline
\hspace{1em} & $0$ & $1$  \\ 
\hline
$0$ & $\cos^2 \theta $ & $0$ \\ 
$1$ & $0$ & $\sin^2 \theta $\\ 
\hline
 \end{tabular}
\end{center}
The mutual information or entropy of entanglement is given by, 
\begin{eqnarray}
I(A;B) & = & S(\rho_{A}) \nonumber \\
& = & -\sin^2\theta\log (\sin^2\theta)-\cos^2\theta\log (\cos^2\theta)
\end{eqnarray}

Next consider the case where the two parties are not cooperating with each other. Suppose Alice choose to measure in the $\sigma_{x}$ basis. She will get the outcomes $+$ or $-$ with equal probability. Bob's state will then be $\ket{\Psi}_{B}(A = \pm) = \ket{0}\cos\theta \pm \ket{1}\sin\theta $. The bipartite state under this setting can be rewritten as,
\begin{eqnarray}
\ket{\Psi_{\theta}} & = & \ket{+}_{A}\bigg(\frac{\ket{0}\cos\theta+\ket{1}\sin\theta}{\sqrt{2}}\bigg)_{B} + \nonumber \\  && \ket{-}_{A}\bigg(\frac{\ket{0}\cos\theta-\ket{1}\sin\theta}{\sqrt{2}}\bigg)_{B}
\end{eqnarray}

The optimal strategy for Bob to discriminate her two states is to measure in the $\sigma_{x}$ basis. Doing so, they end up with the following outcomes,
\begin{center}
\begin{tabular}{| c | c  c |}
\hline
\hspace{1em} & $+$ & $-$  \\ 
\hline
$+$ & $\cfrac{1}{4} (\cos \theta + \sin \theta)^{2}$ & $\cfrac{1}{4} (\cos \theta - \sin \theta)^{2}$ \\ 
$-$ & $\cfrac{1}{4} (\cos \theta - \sin \theta)^{2}$ & $\cfrac{1}{4} (\cos \theta + \sin \theta)^{2}$ \\ 
\hline
 \end{tabular}
\end{center}
The mutual information $I(A;B)$ for this case, as a function of $\theta$, is given by,
\begin{equation}
\begin{split}
I(A;B) = \frac{1}{2}\log (1+\sin 2 \theta )+\log (1- \sin 2\theta) \\
+\frac{\sin 2 \theta}{2}  \bigg(\log (1+\sin 2 \theta)-\log (1- \sin 2\theta)\bigg).
\end{split}
\end{equation}

The mutual information is calculated and compared with entropy of entanglement. From Fig.~\ref{upper}, it is clear that the maximum mutual information between Alice and Bob when they perform disparate measurements and cooperative measurements (entropy of entanglement) are different. In case of disparate measurements, the blue curve corresponds to a measurement strategy that need not be the optimal strategy for Alice, which corresponds to an upper bound to the EA. This result even implies that EA is less than or equal to the entropy of entanglement for pure states.

\begin{figure*}[htp]
\centering
 \subfloat{%
    \xincludegraphics[width=.33\textwidth,label=a)]{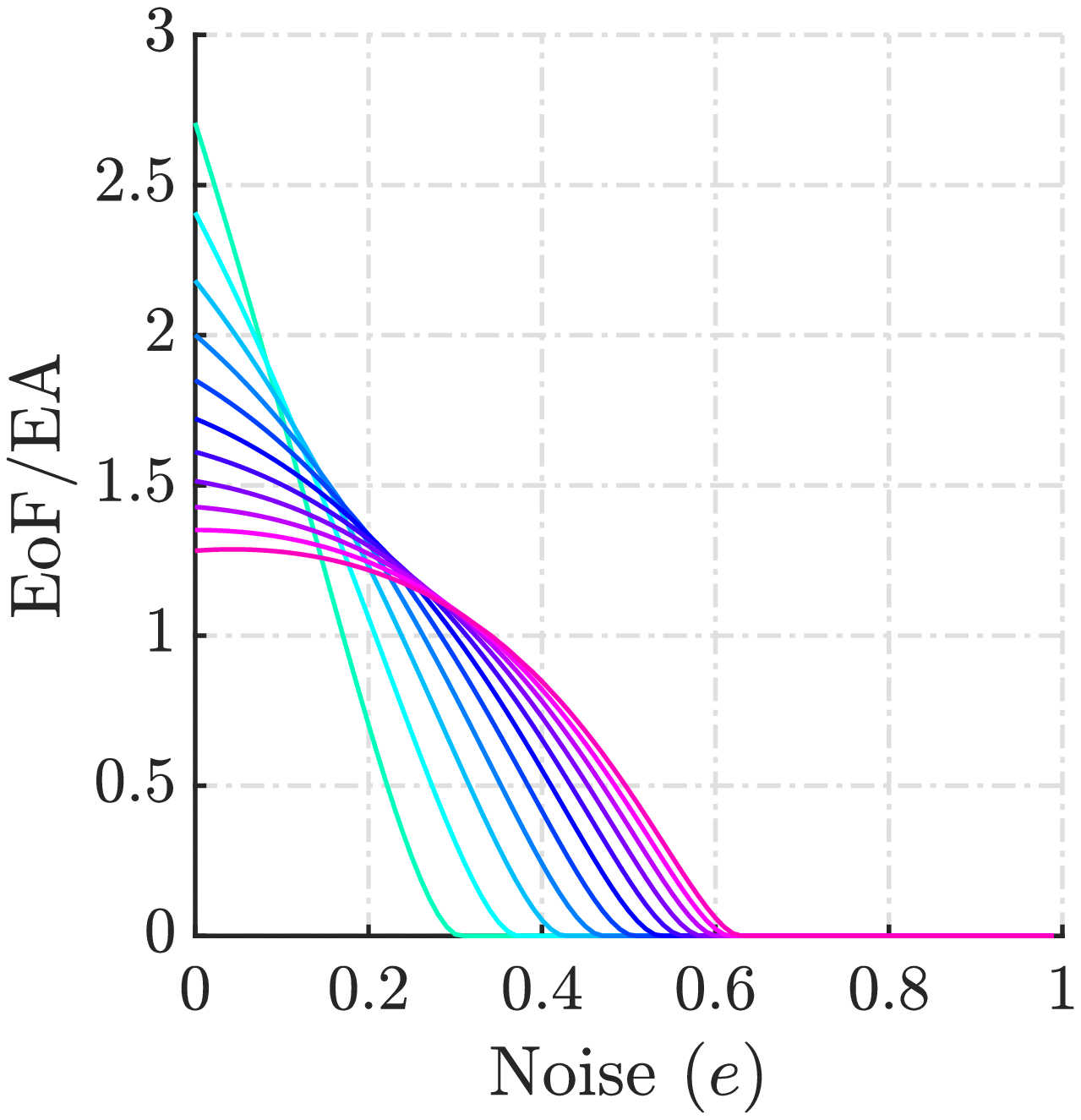}}\hfill
  \subfloat{%
   \xincludegraphics[width=.33\textwidth,label=b)]{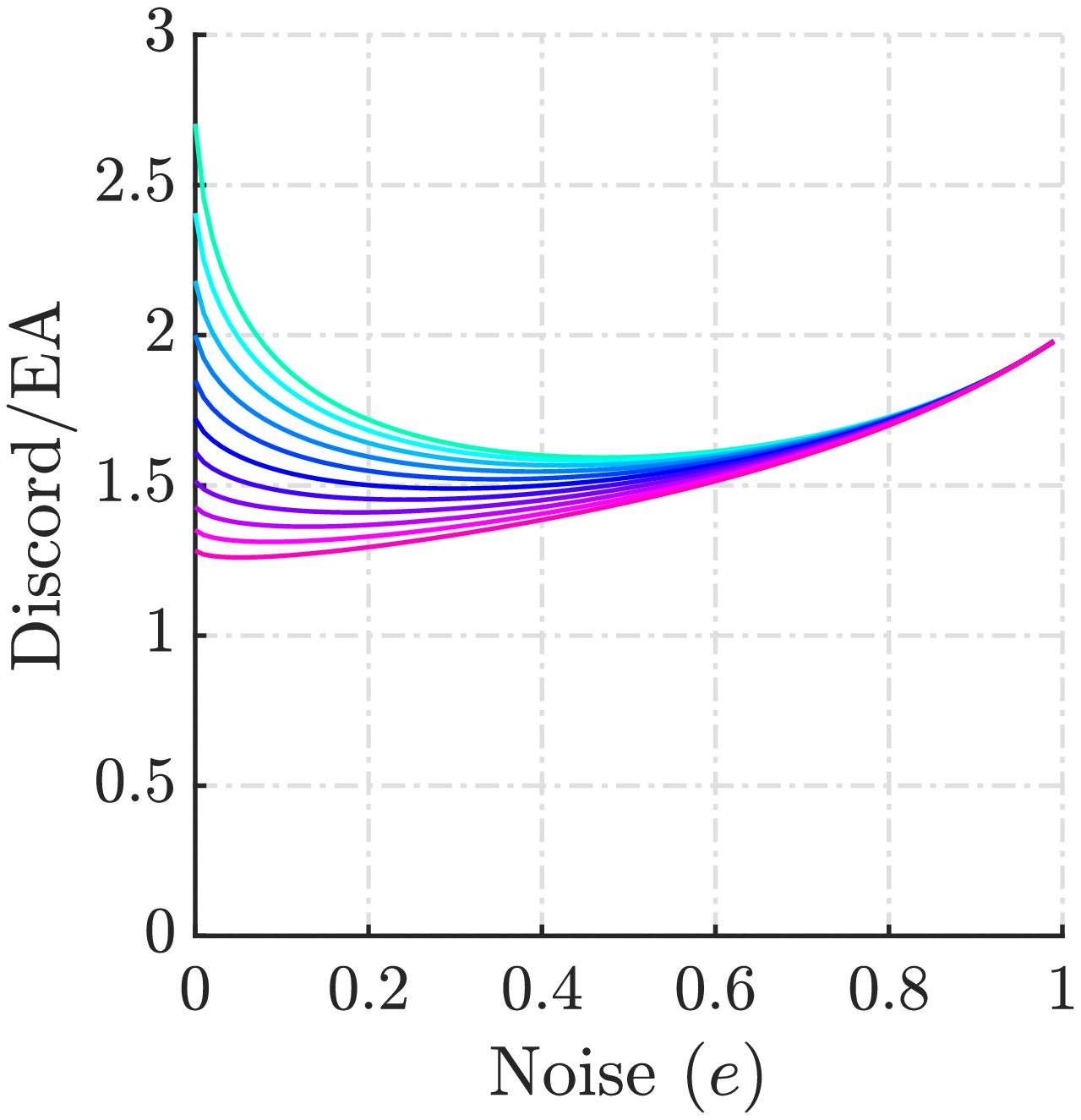}}\hfill
  \subfloat{%
   \xincludegraphics[width=.33\textwidth,label=c)]{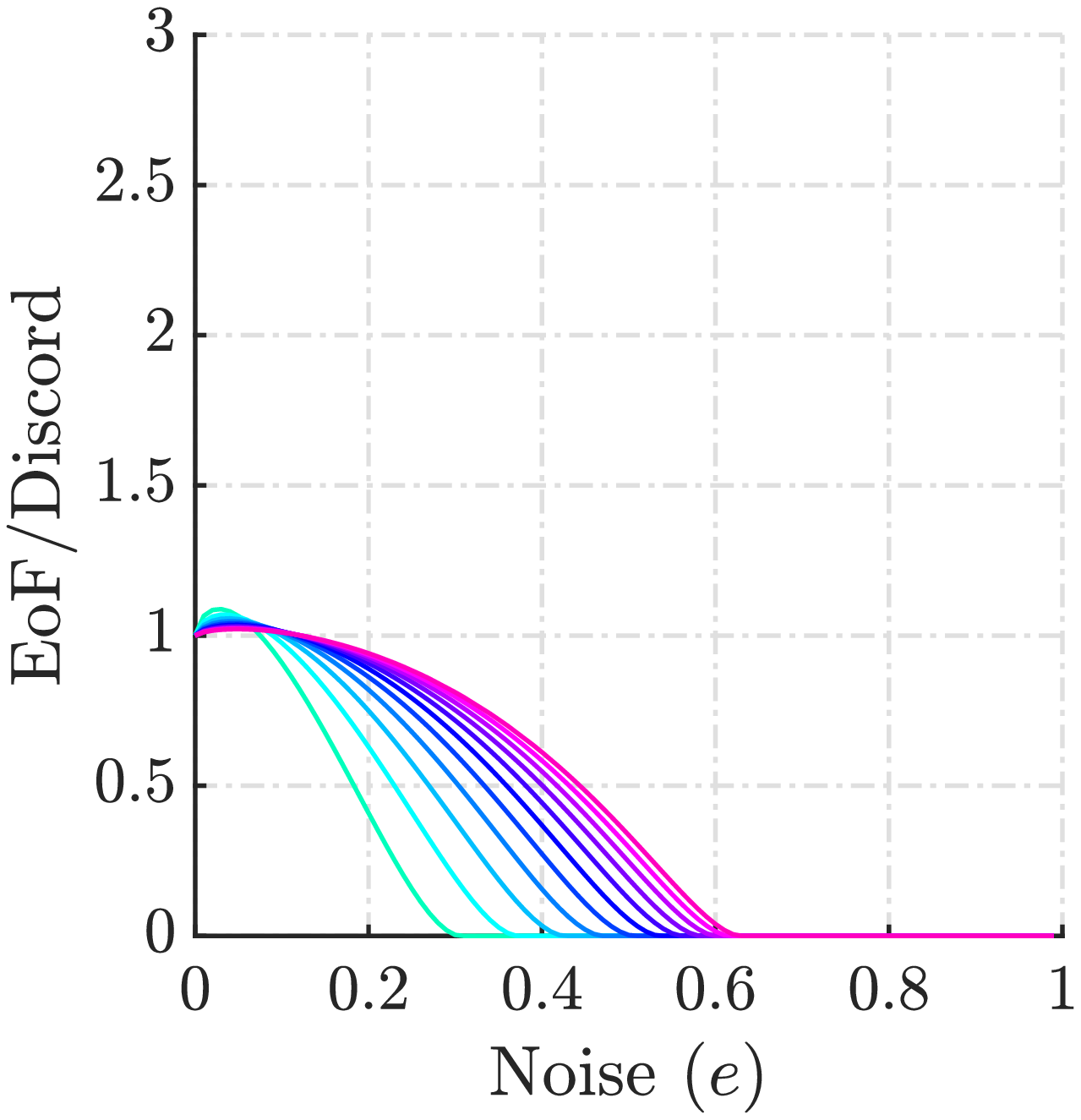}}\\
 \caption{Comparison of EA with EoF and discord for pure states with white noise. Each colour represents how noise affects the correlation measures for a given pure state parametrised by the Schmidt angle~$\theta$. The pink colour represents the pure states that are highly entangled~($\theta \rightarrow \pi/4$) whereas cyan represents states that are not highly entangled~($\theta \rightarrow 0$). At all times EA is less than or equal to discord. Whereas EA is only greater than EoF when the degree of noise present in the mixture is significant.}\label{pure_state_noise_same_noise} 
\end{figure*}

As mentioned earlier, for pure states, quantum discord is a measure of entanglement. However for mixed states, quantum discord can measure correlations beyond entanglement. It is worthwhile looking into how these measures compare when noise is introduced as these class of states are much more relevant from an operational perspective. We compare EA with entanglement of formation which is also an entropy based measure of entanglement. EoF is also interesting because of its intuitive meaning which is relevant from an operational perspective. EoF represents the asymptotic number of standard singlets that is required to locally prepare a state which therefore represents maximum measure of entanglement in a system. Similarly, discord is an entropic measure which has already demonstrated an operational advantage. We compute the EoF for two-qubit states $\rho$ with the following equation,
\begin{equation}
	E(\rho) = h\bigg(\frac{1+\sqrt{1-C^2(\rho)}}{2}\bigg) 
\end{equation}
where $C(\rho)$ is the concurrence and $h(x) = -x\log_{2}x - (1-x)\log_{2}(1-x)$. Concurrence is given by, 
\begin{equation}
C(\rho) \equiv  \text{max} (0,\lambda_1-\lambda_2-\lambda_3-\lambda_4)
\label{concurrenceeqn}
\end{equation}
where $\lambda_1,\lambda_2,\lambda_3,\lambda_4$ are the eigenvalues in decreasing order of the hermitian matrix $R = \sqrt{\sqrt{\rho}\tilde{\rho}\sqrt{\rho}}$ with $\tilde{\rho} = (\sigma_y\otimes\sigma_y)\rho^*(\sigma_y\otimes\sigma_y)$ which is the spin-flipped state of $\rho$ and $\sigma_y$ is a Pauli spin matrix \cite{hill1997entanglement}. Discord was calculated by taking the difference of quantum mutual information and the Holevo quantity $J(\rho,\pi_B)$ \cite{zurek2001,datta2009signatures} maximised over $\pi_B$,
\begin{equation}
D(\rho) =  I(\rho) - \maxi_{\pi_B} \{ J(\rho,\pi_B) \} \label{discordeqn}.
\end{equation}
The Holevo quantity represents the maximum information gained about the system $\rho$ when Bob performs projective measurements $\{\pi_{B}\}$~\cite{holevo2011probabilistic}. EA was numerically calculated using Eq.~\ref{perf} for two-qubit states between Alice and Bob. A minimax optimisation was employed by varying the angle of the rotation Alice and Bob implement before performing their measurement.

In Fig.~\ref{pure_state_noise_same_pure_states}, each colour represents the set of states of the form $\rho = (1-e)(\ket{\Psi_{\theta}} \bra{\Psi_{\theta}}) + e \mathbb{1}/4 $, where $\ket{\Psi_{\theta}} = \ket{00} \cos \theta + \ket{11} \sin \theta$. We can see that as we add white noise ($e$) into the pure state with a given $\theta$, the correlations decrease to eventually become zero. For a given EoF and discord, the numerical results shows that the Werner states bound the maximum amount of EA a pure state with added white noise can have and the lower bound is given by pure states. The results in Fig.~\ref{pure_state_noise_same_pure_states} gives an indication of how the trend varies as the degree of noise present in the system changes. We can see that with sufficient noise the EoF goes to zero while EA is still finite. This signifies that entanglement is a subset of EA. We also note that both EA and discord only goes to zero when the state is maximally mixed. An alternative representation of these results are shown in Fig.~\ref{pure_state_noise_same_noise} which makes the effect of noise more apparent. 

\subsubsection{Bell-diagonal states}
\begin{figure*}
        \center{\includegraphics[width=\textwidth]
        {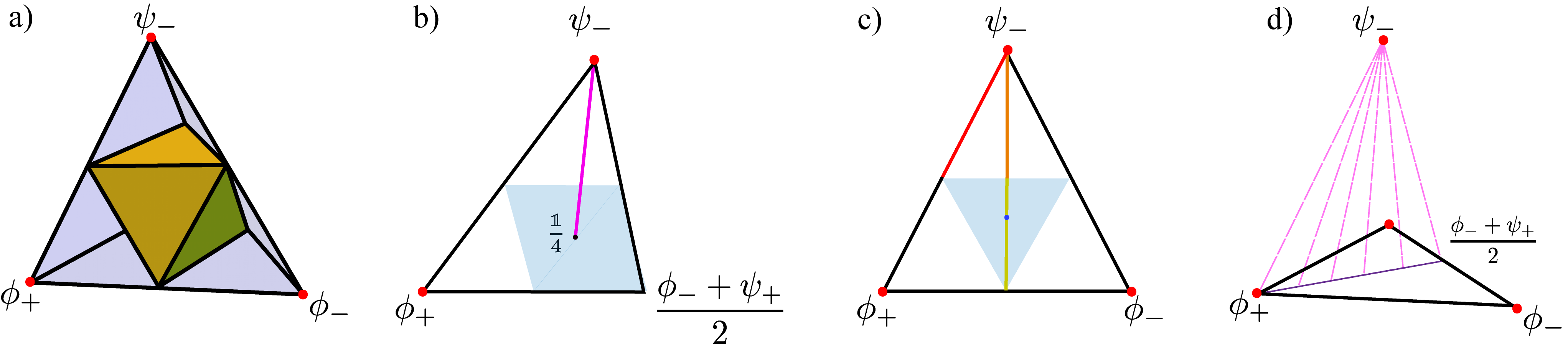}}
        \caption{\label{belld} Bell-diagonal state tetrahedron representation. (a) represents the blue tetrahedron is the entire set of Bell-diagonal states. The Bell states sit at the vertices of the tetrahedron indicated by the red dot. The yellow octahedron is the class of separable states corresponding to the condition $\sum_{j=1}^{3} \abs{c_j} \leq 1$. The region outside the separable states are entangled. (b) and (c) are the Schematic representation of two slices from the Bell-diagonal space. The lines connecting the centre of T and the vertices are maximally entangled states with added white noise. In (b), we represent the special case of Werner states (magenta line) where the maximally entangled state is the Bell singlet. (c) represents one of the faces of the tetrahedron. The face represents the rank 3 states and its edges represent the rank 2 states in the Bell diagonal space. The red line corresponds to the edge states, the orange line represents the type I face states and yellow line, the type II face states. The blue shading represents the separable region on that face. (d) depicts the dotted pink lines that represent the rank 4 Bell states which are the convex combinations of the maximally entangled Bell state $\psi_-$ and face states.}
\end{figure*}

In this subsection, we look at the comparison of EA with entanglement and discord for Bell-diagonal states which are an interesting sub-class of two-qubit states. Bell-diagonal states are three-parameter sets of states that includes maximally entangled states, separable states and classical states which can be represented as points in a tetrahedron. The geometry also produces other regions of interests which will be discussed further in detail. All two-qubit states can be written as, 
\begin{equation}
\rho = \frac{1}{4} (\mathbb{1} \otimes \mathds{1} + \vec{r}.\vec{\sigma} \otimes \mathds{1} + \mathds{1} \otimes \vec{s}.\vec{\tau} + \sum_{ij} c_{ij} \sigma_{i} \otimes \tau_{j}). \label{2qubit}
\end{equation}\\
Here $\mathds{1} $ is the identity operator, $\sigma$ and $\tau$ are the standard Pauli matrices of Alice and Bob respectively. The coefficients $c_{ij}=\mathrm{Tr}(\rho \sigma_i \otimes \tau_j)$ forms a real matrix $C$ which is responsible for the correlation of the state (i.e., joint measurements are necessary to estimate them). The parameters $\vec{r}$ and $\vec{s}$ are local parameters (i.e., only local measurements are necessary to calculate these) and they represent the reduced state. For Bell diagonal states, $\vec{r}$ and $\vec{s}$ are zero and the matrix $C$ is diagonal which implies we only need to consider the values $c_{ij}$ which constitute a vector in a three-dimensional space~\cite{bertlmann2002geometric}. Equation~(\ref{2qubit}) can be rewritten for Bell diagonal states as,
\begin{equation}
\rho_{AB} =  \frac{1}{4} (\mathbb{1}  + \sum_{j=1}^{3} c_{j} \sigma_{j} \otimes \tau_{j}) \label{bell}.
\end{equation}
The bell diagonal states can be pictorially represented as a tetrahedron T (Fig.~\ref{belld}) where the coordinates are the $c_j$'s from Eq.~(\ref{bell}) with vertices (-1,-1,-1), (-1,1,1), (1,-1,1) and (1,1,-1). Within the tetrahedron, the separable states lie inside the octahedron, $|c_1|+|c_2|+|c_3|\leq 1$~\cite{lang2010quantum}. This geometric picture has been used previously to study how non-classical measures change along one-dimensional trajectories traced out by decohering states~\cite{lang2010quantum}. This approach becomes useful in the upcoming analysis.

In addition to the geometrical intuition these states entail, Bell-diagonal states are also relevant because of the fact that any generic two-qubit state can be probabilistically turned into Bell diagonal states by performing local operations~\cite{verstraete2001local,cen2002local}. These operations are probabilistic and the correlation cannot be increased by these operations. As a result, any analysis of these class of states would potentially translate to understanding correlations in two-qubit states as a whole.

\begin{figure*}
	\centering
  	\subfloat{%
    \xincludegraphics[width=.5\textwidth,label=a)]{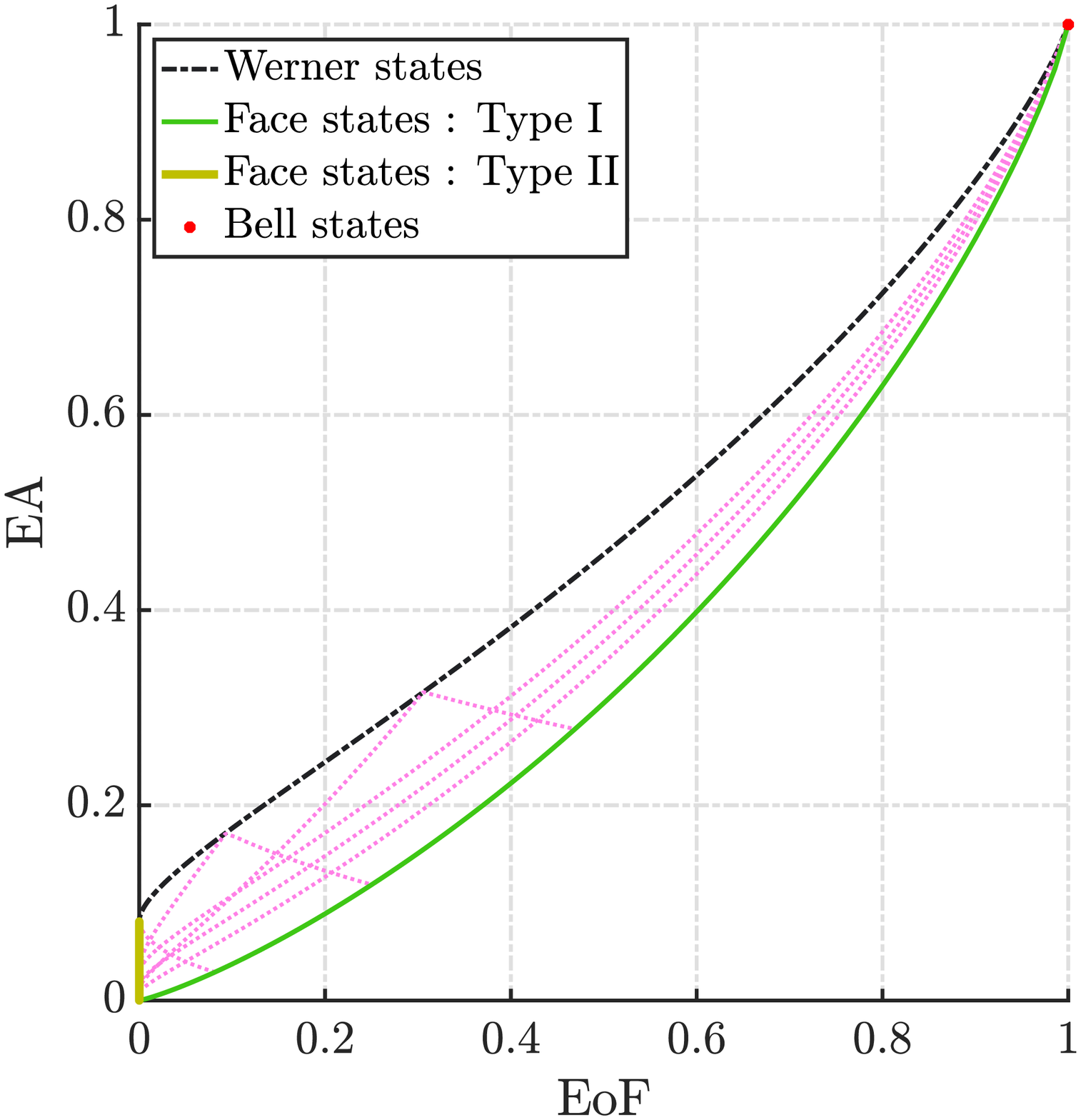}}\hfill
    \subfloat{%
    \xincludegraphics[width=.5\textwidth,label=b)]{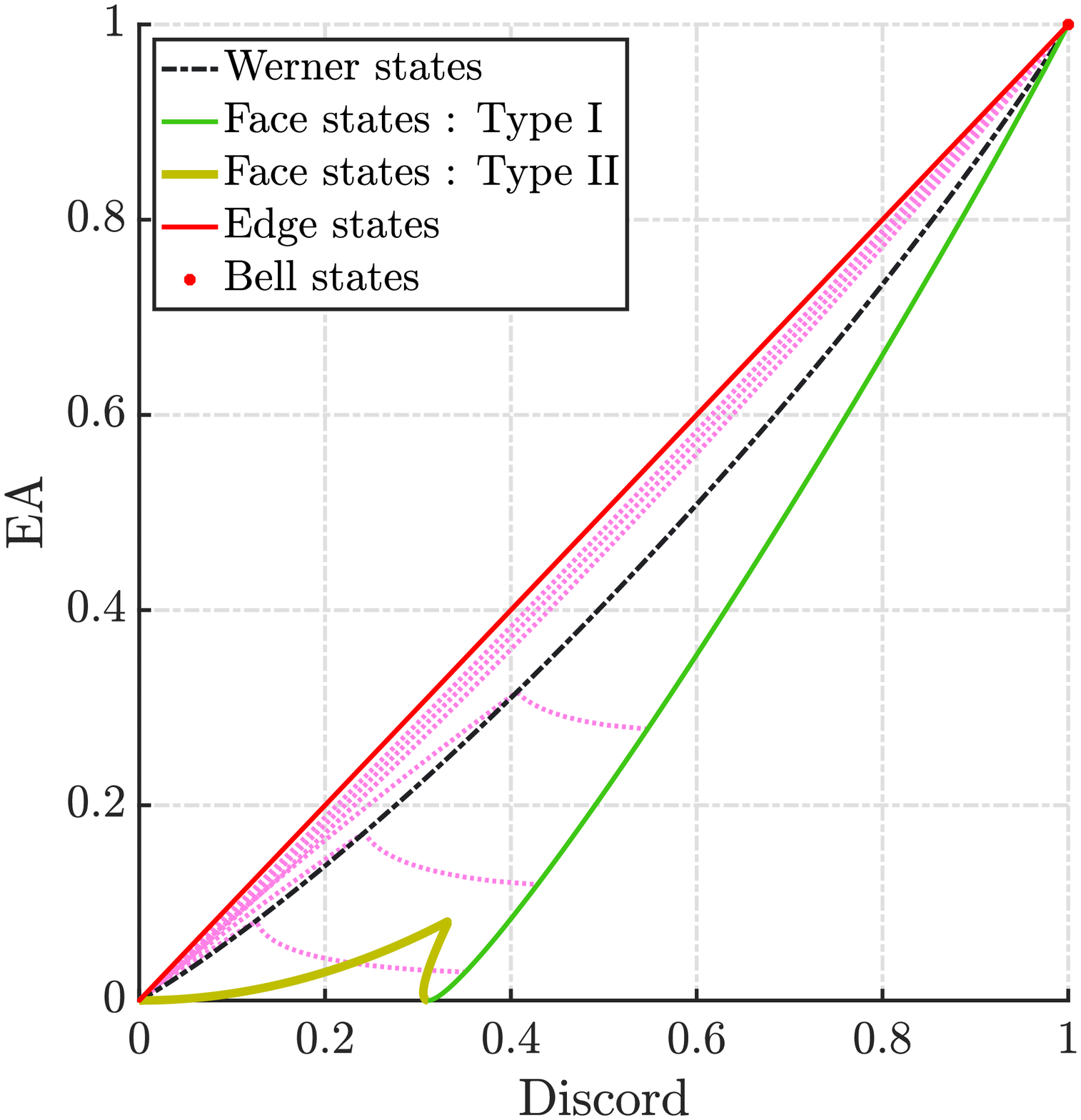}}\\
 	\caption{Comparison of EA with EoF and discord for Bell-diagonal states. The dashed pink lines corresponds to the rank 4 Bell states depicted in Fig.~\ref{belld}(d).}\label{bell_rank} 
\end{figure*}

\begin{figure*}
	\centering
  	\subfloat[EoF]{\xincludegraphics[width=.28\textwidth]{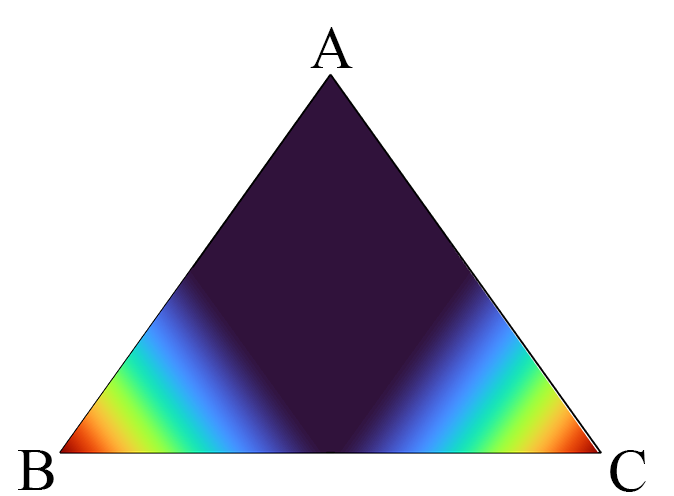}}\hfill
  	\subfloat[EA]{\xincludegraphics[width=.28\textwidth]{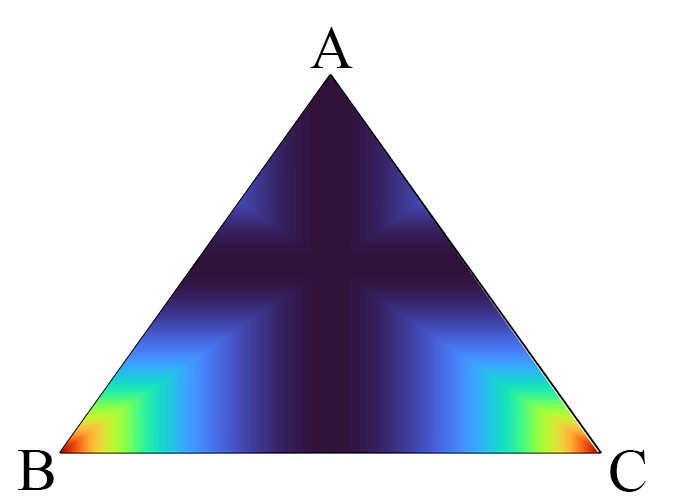}}\hfill
  	\subfloat[Discord]{\xincludegraphics[width=.28\textwidth]{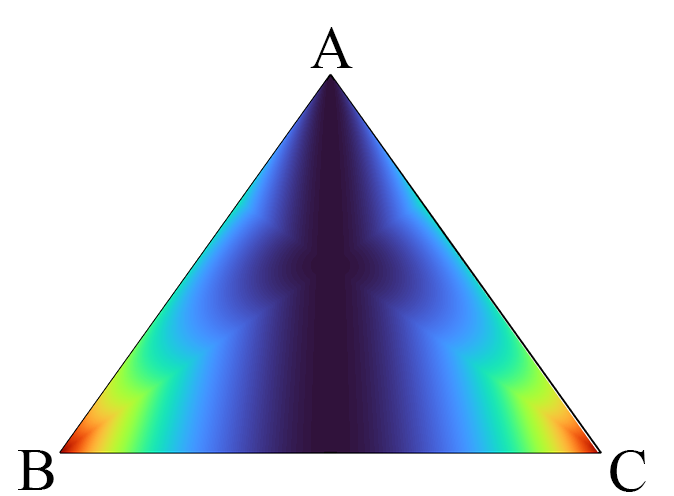}}\hfill
  	\subfloat{\xincludegraphics[width=.055\textwidth]{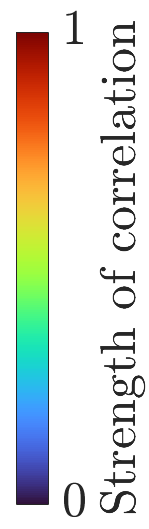}}\\
 	\caption{Strength of correlations on the plane formed by the convex combination of maximally entangled bell states and face states. A is $(\phi_- + \psi_+)/2$, B and C are the other Bell states. The states depicted in Fig.~\ref{belld}(d) are lines along one such plane. Fig.~\ref{bell_face_variation}(a), Fig.~\ref{bell_face_variation}(b) and Fig.~\ref{bell_face_variation}(c) represent EoF, EA and discord respectively. EoF vanishes in the separable region whereas EA and discord are still non-zero for some states in the separable region. The surface plots are also indicative of the hierarchy these measures follow. EoF must be zero when EA is zero, and EA must be zero when discord is zero.}
 	\label{bell_face_variation}
\end{figure*}

\begin{figure*}
	\centering
	\subfloat{\xincludegraphics[width=.308\textwidth,label=a)]{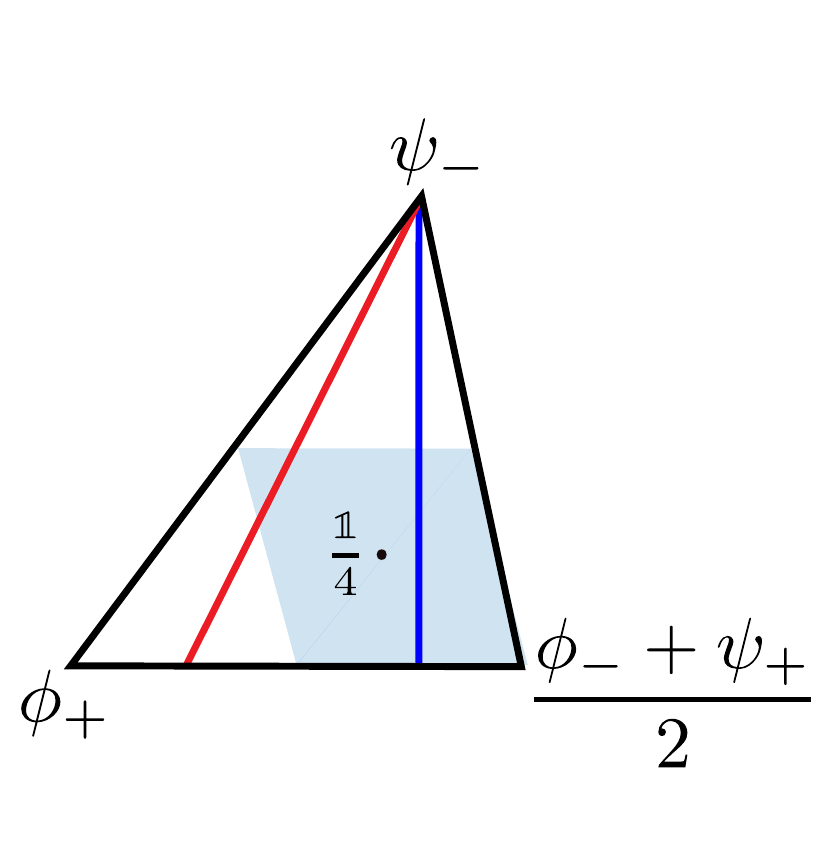}}\hfill
	\subfloat{\xincludegraphics[width=.32\textwidth,label=b)]{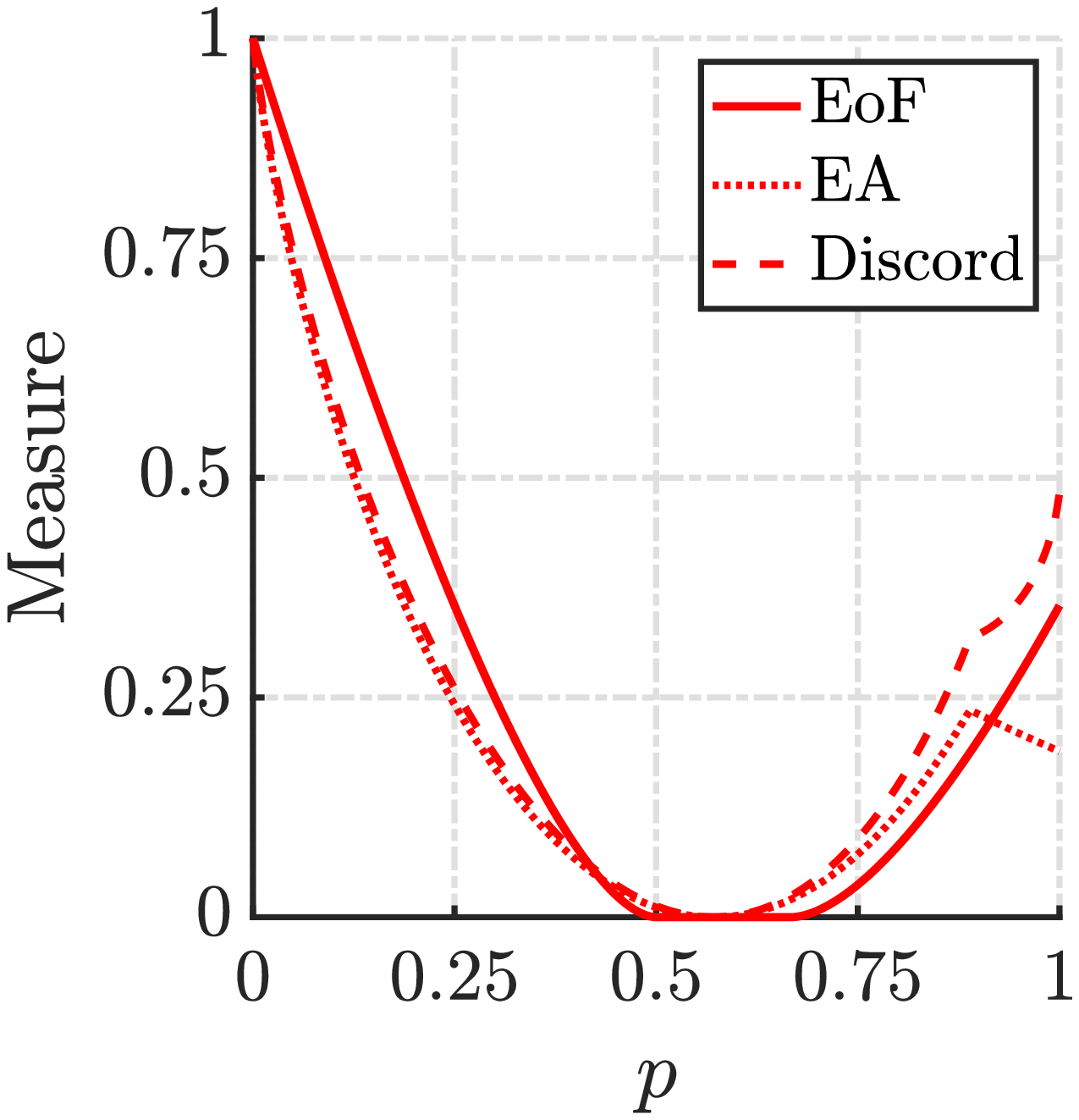}}\hfill
  	\subfloat{\xincludegraphics[width=.32\textwidth,label=c)]{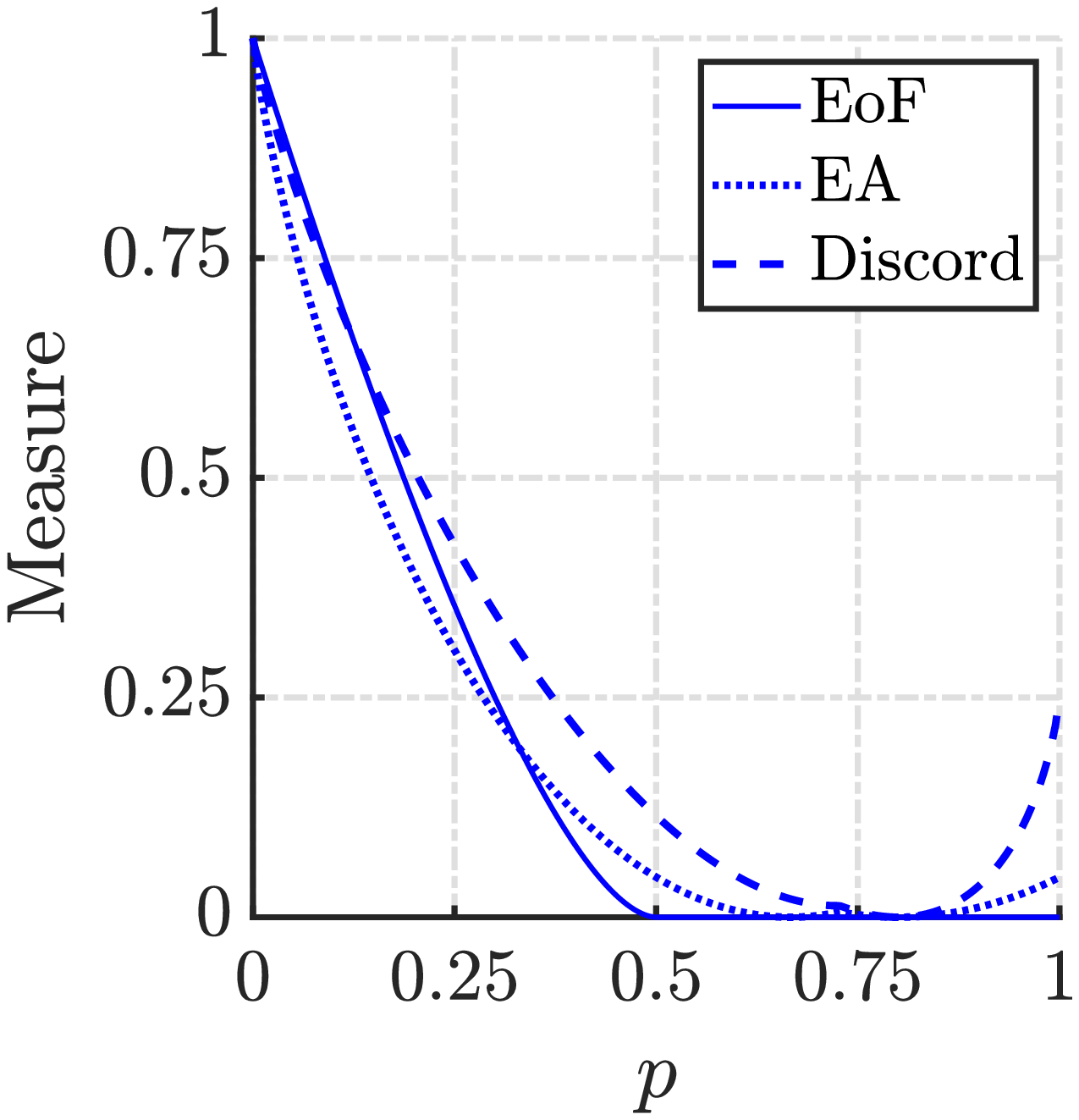}}\\
 	\caption{Variation of correlation measures along the lines in the plane shown in Fig.~\ref{bell_variation}(a). $p$ represents the distance of the state from the vertex $\psi_-$ along the respective lines. The blue shaded region in Fig.~\ref{bell_variation}(a) represents the separable region.}
 	\label{bell_variation}
\end{figure*}

\begin{figure*}
	\centering
  	\subfloat{%
    \xincludegraphics[width=.5\textwidth,label=a)]{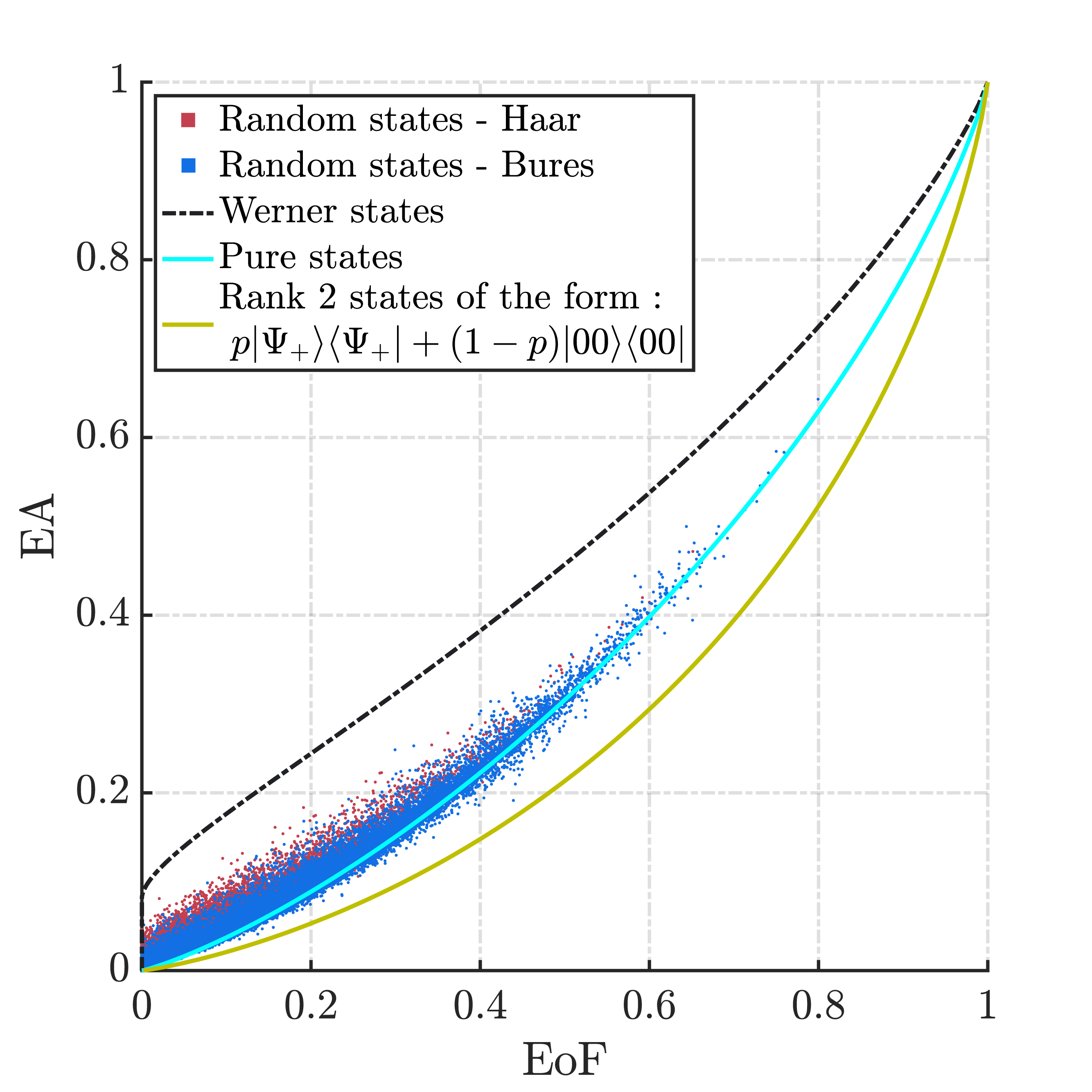}}\hfill
  	\subfloat{%
  	\xincludegraphics[width=.5\textwidth,label=b)]{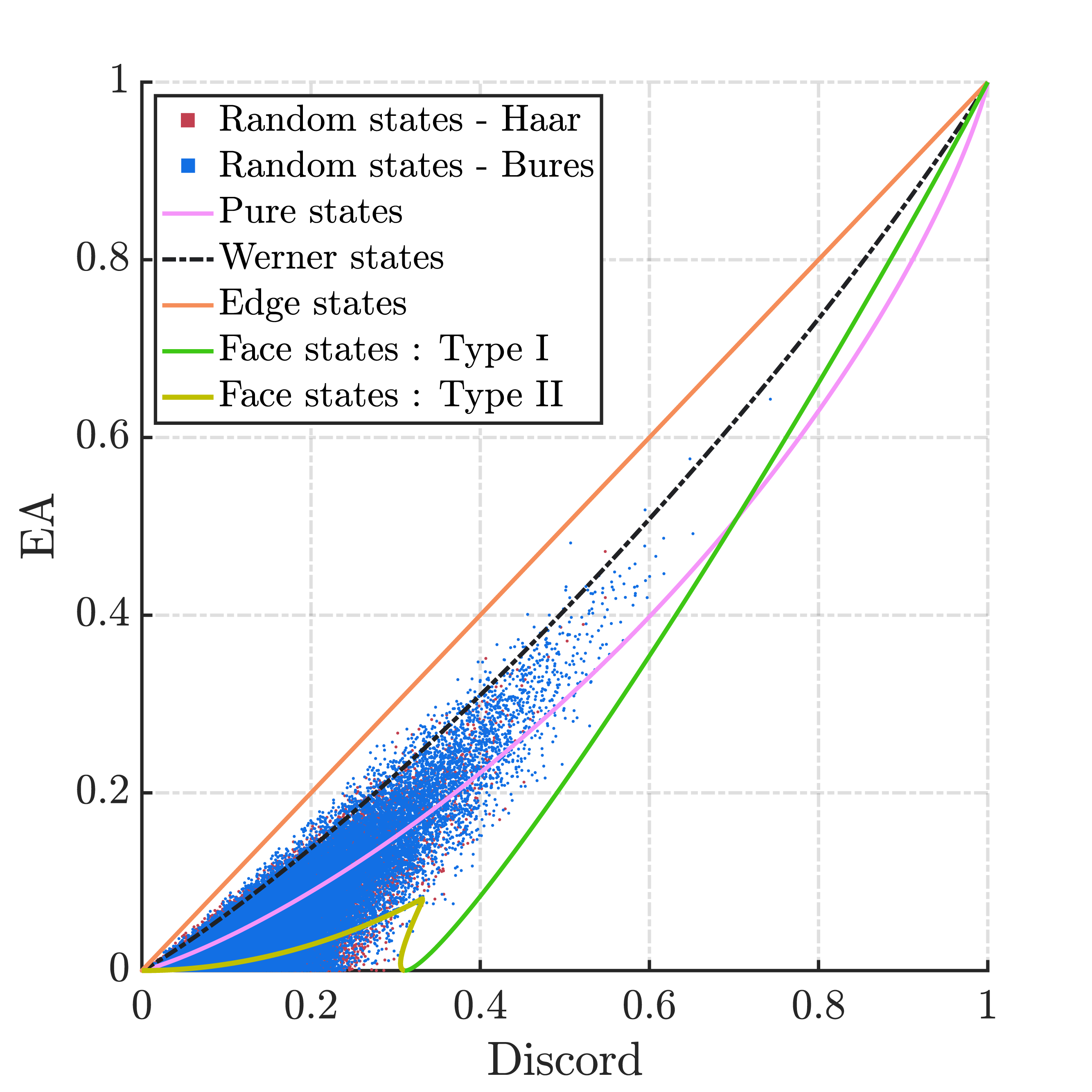}}\\
 	\caption{EA compared with entanglement and discord for $10^5$ randomly generated states.}\label{random_states}
\end{figure*}

By definition, bell diagonal states are convex combination of the four bell states ($\phi^{+}$,$\phi^{-}$,$\psi^{+}$,$\psi^{-}$). Therefore, Eq.~(\ref{bell}) can be rewritten as,
\begin{equation}
\begin{split}
\rho_{AB} & =  p_{1} \ket{\phi^{+}} \bra{\phi^{+}} + p_{2} \ket{\phi^{-}} \bra{\phi^{-}} \\
& + p_{3} \ket{\psi^{+}} \bra{\psi^{+}} + p_{4} \ket{\psi^{-}} \bra{\psi^{-}}\label{density}
\end{split}
\end{equation}
i.e., the eigenstates are four bell states with eigenvalues $p_{1}$, $p_{2}$, $p_{3}$ and $ p_{4}$. This helps in uniformly sampling the entire space of the tetrahedron and systematically looking at distinct groups of states with different ranks. Figure~\ref{belld}(b) and (c) shows two slices of the tetrahedron. By calculating the measures along the different slices of the tetrahedron, we will be able to clearly see how these measures behaves at the point of transition from the inseparable states to separable states (blue shaded region represents the slice of the octahedron on that plane of interest). By definition such separable states will have no entanglement. The positive value of EA even when entanglement is zero would imply EA is a measure that does not rely solely on separability as the only criterion for the presence of quantum correlations.
EA was numerically calculated for all the states shown in Fig.~\ref{belld}. The results of Werner states were consistent with the results seen in the previous subsection where Werner states had the maximum EA for a given EoF when noise is introduced. In general, these are rank 4 states. From Fig.~\ref{bell_rank}, type I face states has the least amount of EA for a given EoF. Also, as expected the rank 3 type II face states have zero EoF and finite EA. The rank 2 edge states have the same trend as that of type I face states. In Fig.~\ref{bell_rank}, we see how these states compare with each other. The dotted pink lines inside the horn are rank 4 states inside the tetrahedron. Each dotted pink line represents how the measure varies along different lines that connects the Bell state vertex and the opposite face (depicted in Fig.~\ref{belld}(d)).

Unlike EA and EoF where the edge states and type I face states have the same value, they have different values for discord. The edge states has the maximum EA for a given discord and one of the edges of the lower boundary is set by type I states~(orange line from Fig.~\ref{belld}(c)). Also the area bounded by the horn in case of EA vs discord is larger than the area in the EA-EoF parametric plot which is expected as some separable states in the Bell-diagonal space would have a finite value of EA and discord. Fig.~\ref{bell_face_variation} represents the variation of the three correlations on the plane which contain the pink dotted lines from Fig.~\ref{belld}(d) and Fig.~\ref{bell_variation} represents two lines along this plane where $p$ represents the distance of the point from the vertex $\psi_-$. From Fig.~\ref{bell_variation}(a) and (b), we see the EoF and EA decreases gradually as you move through the line which starts from one of the vertices and ends at the opposite face. As the point enters the separable region, the EoF goes to zero while EA and discord remains finite. As the line exits the separable region, EoF increases again while EA and discord increases however with an exception. There is a kink in the red lines in Fig.~\ref{bell_variation}(b) representing the EA and discord. Past the kink, discord and EoF are increasing while EA is decreasing. This not only adds to how EA differs from EoF but also points out how EA and discord behave differently particularly in the separable region. We can also say the three measures vanish to zero and gain correlations at a different rate which is consistent with the expectations regarding the hierarchy. This is evident from Fig.~\ref{bell_face_variation}. From this, we expect a state with zero EA to have zero EoF which is indeed the trend we see from the results in Fig.~\ref{bell_rank}(a). We can see that the states are grouped more towards the EA axis as EoF is decreasing. Similarly, we can see EA can be zero even when discord is non-zero.

It is worthwhile pointing out that the results in Fig.~\ref{bell_rank} do not represent the complete set of states present in the Bell-diagonal states. However because of the geometry and how states of similar properties are confined in a specific region, the results are indicative of properties of all bell-diagonal states. In addition we also know which group of states bounds the values that makes this analysis sufficient, i.e., a larger number of points would only fill up the horn rather than showing any trends outside the region shown. This was verified numerically. The Bell-diagonal picture significantly helped in understanding how these correlations behave as one translates through several points in the tetrahedron T. The geometry and its implications helps to compare states of different ranks easily.

As pointed out earlier, it is true that an arbitrary two-qubit state can be probabilistically converted to the Bell diagonal state. This means that the trends which are evident from Fig.~\ref{bell_rank} are indicative of a trend that would be present for every two-qubit states. To look at how these measures compare among each other, the results from Fig.~\ref{bell_rank} can be extended to the entire two-qubit space by evaluating these measures for randomly generated two-qubit states which will represent a more generalised comparison. The random arbitrary two-qubit states were generated uniformly according to the Haar and Bures measures~\cite{zyczkowski2011generating}. These results are shown in Fig.~\ref{random_states}. We could see a similar trend to what we saw with the Bell-diagonal states where the points are more grouped towards the EA axis for the EA-EoF parametric plot and grouped more towards the discord axis for the EA-discord plot. This is expected as a consequence of the hierarchy proven earlier.

Interestingly for the EA-discord comparison, there were no points outside the region formed by envelope of the type I Bell-diagonal face states and the pure states of the form from Eq.~(\ref{pure_eqn}) and Bell-diagonal edge states, whereas for the EA-EoF parametric plot, there were few states outside the horn formed by envelope of these states. However numerical calculations showed that there exist no states outside the contour formed by rank 2 states of the form $\rho(p) = p\ket{\psi_+}\bra{\psi_+} + (1-p)\ket{00}\bra{00}$. In brief, the lower boundary for the parametric relation of EA-EoF would be formed by $\rho(p)$ and for the EA-discord would be jointly formed by the envelope of face states and the pure states from Fig.~\ref{random_states}(b). These results are consistent with the previous trend and indicate no anomalies for the proposed hierarchy. 

These differences presented with the supporting results from subsections (1), (2) and (3) points out how EA quantitatively differs from entanglement and discord when measuring mixed state correlations. More importantly EA also differs from entanglement for pure states~(discord measures entanglement in this case).

\section{Discussion}

EoF and discord rely on the cooperative nature of the correlated subsystems. EA being a nonclassical correlation measure that is based on an adversarial nature, probes a distinct aspect of quantum correlations. In this paper, we show direct evidence of how EA does not converge to other measures of entanglement for pure states. This property is in clear contrast to discord, which converges to EoF. Because of the stark adversarial roles Alice and Bob play, EA is more sensitive to noise compared to discord. The adversarial nature of the correlations is better revealed by EA when the states are mixed. The difference between the three analysed measures becomes more obvious for mixed states and their hierarchy becomes more evident when the amount of noise increases. EA, therefore, offers instead an intuitive understanding of mixed state correlations otherwise not probed, based on measurements on the subsystems. However, in this work, we limited the discussion to two-qubit states where the subsystems are represented in equal dimensions. It will be interesting to look into the nature of EA when one of the subsystems is represented in a larger Hilbert space. Further, EA can be extended to higher-dimensional systems and it is intriguing to check whether EA has any operational interpretation as well.

\section*{Data availability}
The data that support the findings of this study are available from the corresponding author upon reasonable request.

\section*{Code availability}
The codes that support the findings of this study are available from the corresponding author upon reasonable request.

\bibliography{ref.bib}

\section*{Acknowledgements}
This research was funded by the Australian Research Council Centre of Excellence for Quantum Computation and Communication Technology (Grant No. CE110001027). P.K.L. acknowledges support from the ARC Laureate Fellowship FL150100019.

\section*{Author contributions}
S.A. conceived the project. S.A. and B.S. developed the theory and performed the numerical analysis. B.S. and S.A. wrote the manuscript. All authors contributed to discussions regarding the
results in this paper. P.K.L. supervised the project.

\section*{Competing Interests}
The authors declare no competing financial or non-financial interests.

\end{document}